\def\rfr#1{eq. (\ref{#1})}

\def\dert#1#2{\frac{{{d}}{#1}}{{{d}}{#2}}}              

\def\eqi{\begin{equation}}
\def\eqf{\end{equation}}
\def\eqia{\begin{eqnarray}}
\def\eqfa{\end{eqnarray}}
\def\rp#1#2{{#1\over#2}} \def\lb#1{\label{#1}}

\def\bds#1{\boldsymbol{#1}}

\documentclass{aastex}
\usepackage{spr-astr-addons}
\usepackage{url}
\usepackage{amsmath,amsthm,amscd,amssymb}
\usepackage{latexsym}
\usepackage{graphicx,epsfig}

\RequirePackage{color}

\newcommand{\emaila}{lorenzo.iorio@libero.it}

\begin{document}

\title{Classical and relativistic node precessional effects in WASP-33b and perspectives for detecting  them}
\shortauthors{L. Iorio}

\author{Lorenzo Iorio\altaffilmark{1} }
\affil{M.I.U.R., F.R.A.S.: Viale Unit\`{a} di Italia 68, 70125, Bari (BA), Italy.}

\email{\emaila}

\begin{abstract}
WASP-33 is a fast rotating, main sequence star which hosts a hot Jupiter moving along a retrograde and almost polar orbit with semi-major axis $a=0.02$ au and eccentricity provisionally set to $e=0$. The quadrupole mass moment $J_2^{\star}$ and the proper angular momentum $S_{\star}$ of the star are 1900 and 400 times, respectively, larger than those of the Sun. Thus, huge classical and general relativistic non-Keplerian orbital effects should take place in such a system. In particular, the large inclination $\Psi_{\star}$ of the orbit of WASP-33b to the star's equator allows to consider the node precession $\dot\Omega$ and  the related time variation $d t_d /dt$ of the transit duration $ t_d $. The WASP-33b node rate due to $J_2^{\star}$ is $9\times 10^9$ times larger than the same effect for Mercury induced by the Sun's oblateness, while the general relativistic gravitomagnetic node precession is $3\times 10^5$ times larger than the Lense-Thirring effect for Mercury due to the Sun's rotation. We also consider the effect of the centrifugal oblateness  of the planet itself and of a putative distant third body X. The magnitudes of the induced time change in the transit duration are of the order of $3\times 10^{-6}, 2\times 10^{-7}, 8\times 10^{-9}$  for $J_2^{\star}$, the planet's rotational oblateness and general relativity, respectively. A yet undiscovered  planet X with the mass of Jupiter orbiting at more than 1 au would induce a transit duration variation of less than $4\times 10^{-9}$. A conservative evaluation of the accuracy in measuring $d t_d /dt$ over 10 yr points towards $\approx 10^{-8}$. The analysis presented here will be applicable also to other exoplanets with similar features if and when they will ne discovered.
 \end{abstract}


\section{Introduction}
HD15082 (WASP-33) is an early type, A5mA8F4 star \citep{Gren}, located at 116 pc from us \citep{Perry}, which has the peculiarity of hosting a close hot Jupiter moving along a 1.22 d retrograde  orbit highly inclined with respect to the stellar equator \citep{Wasp33}. Other planets with the same characteristics have been recently discovered: WASP-2b \citep{wasp2}, WASP-5b \citep{wasp5,wasp5b}, WASP-8b \citep{wasp8}, WASP-15b \citep{wasp15}, WASP-17b \citep{Ross} and HAT-P-7b \citep{hat1,hat2}. Such orbital features are unexpected since the stars and the planets usually originate from the same disk. A possible explanation may be the presence of yet undiscovered planets which have acted in the past in such a way to induce the misalignment of the systems' angular momenta \citep{Ross,Wasp33}. Evidence for a distant third body in the HAT-P-7 system has been, actually, found \citep{hat2}; also WASP-5 may host a third body \citep{wasp5}.

WASP-33 rotates rapidly, almost $1-2$ orders of magnitude faster than the other stars hosting planets moving along highly inclined orbits, so that it is expected that its shape presents relatively large deviations from sphericity. Indeed, as we will show, its quadrupolar mass moment $J^{\star}_2$ should be 1900 times larger than that of the Sun, while a reasonable evaluation of the magnitude of its proper angular momentum $S_{\star}$ points toward a value about 400 times larger than the solar one. Such figures, in addition to the small orbit of its planet, having a semi-major axis $a=0.02555$ au \citep{Wasp33}, suggest that huge non-Keplerian orbital effects should take place in the WASP-33 system. They are larger than those occurring in the other similar planetary systems because of their wider orbits\footnote{Actually, WASP-5 harbors a planet with semi-major axis $a=0.027$ au \citep{wasp5}, but the rotation of the star is 21 times slower than that of WASP-33.} ($a=0.032-0.079$ au). Such effects could be fruitfully used to determine or constrain some key stellar and/or planetary parameters. To this aim, let us note that if, on the one hand, the periastron $\omega$ cannot be used because of the provisionally assumed circularity of the orbit of WASP-33b \citep{Wasp33}, on the other hand its high inclination to the stellar equator makes feasible, at least in principle, the use of the node $\Omega$ which may translate into observable effects of the photometry of the system like, e.g., the transit duration \citep{Miralda}.

Among the dynamical effects considered, we will examine also the impact of the general relativistic gravitomagnetic field \citep{Mash}, induced by bodies' rotation, which in the test particle limit yields the Lense-Thirring effect \citep{LT}. Indeed, it is the only general relativistic effect affecting the node of an orbiting body, and its expected precession for WASP-33b is $3\times 10^5$ times larger than the Lense-Thirring node rate of Mercury in the field of the Sun. Moreover, as we have noted before, it is likely that  WASP-33b is not the only planet hosted by HD15082; if  other distant bodies will be actually discovered, it will be crucial reconstructing the dynamical history of the entire planetary system. In doing that, even small accelerations like the gravitomagnetic one, although $4\times 10^{-8}$ times smaller than the dominant Newtonian monopole, may play a non negligible role. Such a view is supported by the fact that the inclusion of general relativity\footnote{Actually, in this case the general relativistic terms considered are those which depend only on the masses of the bodies, dubbed gravitoelectric \citep{Mash}, not the gravitomagnetic ones. } in the latest long-term numerical integrations of the equations of motion of the solar system \citep{Lask} has drastically changed the probability of the occurrence of a secular resonance between Mercury and Jupiter, which would induce a fatal increase of the eccentricity of the orbit of Mercury, from some tens percent to  $1\%$; the general relativistic accelerations of Mercury and Jupiter are just $7\times 10^{-7}$ and $5\times 10^{-9}$ of the corresponding Newtonian monopoles, respectively. Last but not least, a gravitomagnetic effect may alias the signature of a yet undiscovered further planet X for certain values of its mass and distance.

The influence of some rotation-independent, gravitoelectric effects of general relativity in some extrasolar planets has been treated in \citet{Miralda,Adamsa,Adamsb,Adamsc,Iorio,Heyl,Jordan,Pal,Ragozzine,Li}. In particular, the perspectives in measuring the gravitoelectric correction to the third Kepler law have been considered by \citet{Iorio} and \citet{Ragozzine}. \citet{Miralda}, \citet{Heyl}, \citet{Jordan}, \citet{Pal} and \citet{Ragozzine} dealt with the possibility of detecting the gravitoelectric periastron precessions, while \citet{Li} looked at the gravitoelectric secular change of the mean anomaly connected to the variation of the periastron time transit $t_p$. \citet{Adamsa,Adamsb,Adamsc} studied the impact of the general relativistic gravitoelectric terms on the long-term, secular interactions among multiple planetary systems.

The paper is organized as follows. In Section \ref{una} we set the scene by evaluating some physical parameters of the star and the planet which  enter the dynamical effects analyzed subsequently. In Section \ref{due} we work out the node precessions due to the oblateness of the star and the planet, a putative third body X, and general relativity.
Section \ref{quattro} summarizes our conclusions.
\section{Setting the scene: the relevant parameters of the system}\lb{una}
In Table \ref{stella} we quote the relevant parameters of the host star.
\begin{table*}[t]
\caption{Relevant stellar parameters \citep{Wasp33}. We used $M_{\odot}=1.98895\times 10^{30}$ kg, $R_{\odot}=695.99\times 10^6$ m. In deriving the star's angular speed $\Xi_{\star} = u_{\star}/(R_{\star}\sin I_{\star} )$, with $u_{\star}\doteq v_{\star}\sin I_{\star}$, see the discussion in the text. The value displayed for the tidal parameter $q^{\star}_t$ is an upper bound corresponding to \citep{Wasp33} $m^{\rm p}_{\rm max}=4.1\ m_{\rm Jup}$ for the planet's mass $m_{\rm p}$. The value of the relative semi-major axis used in $q_t^{\star}$ is $a=0.02555\pm 0.00017$ au \citep{Wasp33}. The figure for $J^{\star}_2$ has been obtained by using \citep{Claret} $k_2^{\star}=0.03$ for the star's Love number. The moment of inertia factor is defined as $\alpha_{\star}\doteq C_{\star}/M_{\star} R^{\star 2}_{\rm eq}$, so that the star's proper angular momentum is $S_{\star}=\alpha_{\star} M_{\star}R^{\star 2}_{\rm eq}\Xi_{\star}$. The value used for the flattening $f_{\star}$ is the average of $f^{\star}_{\rm min}$ and $f^{\star}_{\rm max}$.  }\label{stella}
\begin{tabular}{@{}lll}
\hline
Parameter (units) &  Numerical value  \\
\hline
$M_{\star}\ (M_{\odot})$ & $1.495\pm 0.031$\\
$R_{\star}\ (R_{\odot})$ & $1.444\pm 0.034$ \\
$u_{\star}$ (km s$^{-1}$)& $86.48\pm 0.06$\\
\hline
$\Xi_{\star}$ (s$^{-1}$) & $8.6\times 10^{-5}$ \\
$q^{\star}_r$ & $0.0379461$\\
$q^{\star}_t$ & $-1.42772\times 10^{-4}$\\
$J^{\star}_2$ & $3.80175\times 10^{-4}$ \\
$f_{\star}$ & $0.0302587$ \\
$\alpha_{\star}$ & $0.277011$ \\
$S_{\star}$ (kg m$^2$ s$^{-1}$) & $7.16483\times 10^{43}$\\
\hline
\end{tabular}
\end{table*}
A careful discussion is required about the value  of the star's angular speed $\Xi_{\star}$, evaluated from
 \eqi \Xi_{\star}=\rp{u_{\star}}{R_{\star}\sin I_{\star}}\lb{angula},\eqf quoted in Table \ref{stella}. In \rfr{angula} $u_{\star}\doteq v_{\star}\sin I_{\star}=86.48\pm 0.06$ km s$^{-1}$ \citep{Wasp33}, and
$I_{\star}$ is the angle between  the stellar spin axis and the line of sight, not to be confused with the angle $i=87.67\pm 1.81$ deg \citep{Wasp33} between the orbital angular momentum and the line of sight. Both such angles are defined in such a way that  $I_{\star}=i=0$ when the angular momenta and the line of sight, oriented towards the observer, are parallel, and $I_{\star}=i=\pi$  when they are antiparallel. Actually, $I_{\star}$ is unknown; a direct determination of it may be possible by measuring and interpreting asteroseismological oscillations \citep{Gizon},  photometric modulations produced by starspots\footnote{The Kepler mission would be able to accomplish such tasks, but, unfortunately, it cannot see WASP-33. Anyway, time has been awarded on the MOST satellite to analyze the transits,
search for eclipses and characterize the pulsation spectrum of the
host star. (Collier Cameron A., private communication, June 2010).} \citep{Henry}, and polarized light curves \citep{polar}.
Anyway, some reasonable, a-priori constraints can be posed on $I_{\star}$.
 First of all, the natural condition
 \eqi \sin I_{\star}-\rp{u_{\star}}{c}>0\eqf must be satisfied in order to keep the linear rotation speed $v_{\star}$ of the star smaller than the speed of light $c$.
 It implies a preliminary range of admissible values
 \begin{equation}
\begin{array}{lll}
\overline{I^{\star}}_{\rm min}&=&0.016\ {\rm deg},\\ \\
\overline{I^{\star}}_{\rm max}&=&179.983\ {\rm deg}.
\end{array}\lb{angoliluce}
 \end{equation}
In fact,  tighter constraints  on $I_{\star}$ come from the condition that the centrifugal acceleration at the equator must be smaller than the gravitational acceleration at the star's surface, i.e. it must be
\eqi \sin I_{\star}>u_{\star}\sqrt{\rp{R_{\star}}{GM_{\star}}}.\eqf
It implies
 \begin{equation}
\begin{array}{lll}
{I}^{\star}_{\rm min}&=&11.22\ {\rm deg},\\ \\
{I}^{\star}_{\rm max}&=&168.77\ {\rm deg}.
\end{array}\lb{minimoa}
 \end{equation}
By numerically investigating the star's angular speed $\Xi_{\star}$ as a function of $I_{\star}$ according to \rfr{angula} within the range of \rfr{minimoa},
%
%
%
%
it turns out that
\begin{equation}
\begin{array}{lll}
{\Xi}^{\star}_{\rm min}&=& 8.6\times 10^{-5}\ {\rm s}^{-1},\\ \\
{\Xi}^{\star}_{\rm max}&=& 4.5\times 10^{-4}\ {\rm s}^{-1},
\end{array}\lb{minimo}
 \end{equation}
so that
\eqi \rp{\Xi^{\star}_{\rm max}}{\Xi^{\star}_{\rm min}}=5.2.\eqf
The minimum occurs for $\tilde{I_{\star}}=90$ deg.
Anyway, it results that $\Xi_{\star}$ remains below $1.5\times 10^{-4}$ s$^{-1}$  in about  $70\%$ of the allowed range for $I_{\star}$; thus, we conclude that $\Xi_{\rm min}^{\star}$ is somewhat representative of the most likely values for $\Xi$.
In fact, such an argument can be made even more plausible and stringent by numerically investigating the ratio of the stellar centrifugal acceleration to the gravitational one as a function of the allowed values of $I_{\star}$.
%
%
%
%
It turns out that the tails in which $\Xi_{\star}>1.5\times 10^{-4}$ s$^{-1}$ correspond to values of the star's centrifugal acceleration larger than $20\%$ of the gravitational pull. Such figures are clearly highly unrealistic in view of the unobserved associated surface phenomena which, instead, would take place.
Just for a comparison, $A^{\odot}_{\rm cen}/A^{\odot}_{\rm grav}\approx 10^{-5}$ for the Sun. It results that for $I_{\star}=\tilde{I_{\star}}$
the star's centrifugal acceleration reduces to about $4\%$ of the gravitational one, which is its minimum; the same substantially holds for about $60\ {\rm deg}\lesssim I_{\star}\lesssim 120\ {\rm deg}$ in which $\Xi_{\star}/\Xi^{\star}_{\rm min}\approx 1.1$.
The previous arguments motivate our choice of $\Xi_{\rm min}^{\star}$ for the stellar angular speed in Table \ref{stella} and in the rest of the paper.

Concerning the first even zonal harmonic $J^{\star}_2$, it has been computed as \citep{Ragozzine}
\eqi J^{\star}_2\doteq\rp{k_2^{\star}}{3}\left(q^{\star}_r-\rp{q^{\star}_t}{2}\right),\eqf
where\footnote{Actually, in $q_t^{\star}$ the relative star-planet distance $r$, which, in general, is not constant due to the eccentricity $e$, is present. In this case, we have replaced it with the relative semi-major axis $a$ because $e=0$.} \citep{Ragozzine}
\eqi q^{\star}_r\doteq \rp{\Xi_{\star}^2 R_{\star}^3}{GM_{\star}},\ q^{\star}_t\doteq-3\left(\rp{R_{\star}}{a}\right)^3\left(\rp{m_{\rm p}}{M_{\star}}\right)\eqf are related to the rotational and tidal potentials of the star,
and \citep{Claret} $k_2^{\star}\approx 0.03$ for main sequence Sun-like stars.   From Table \ref{stella} it results that the contribution to $J^{\star}_2$ of the tidal distortion raised by the planet of mass $m_{\rm p}$ on the star is negligible.
 It is interesting to note that the value of Table \ref{stella} for $J^{\star}_2$ is about $1900$ times larger than that of the Sun, which is of the order of $J_2^{\odot}=2\times 10^{-7}$ \citep{J2Sun}.

 In regards as the flattening, defined as \eqi f_{\star}\doteq \rp{R^{\star}_{\rm eq}-R^{\star}_{\rm pol}}{R^{\star}_{\rm eq}},\eqf it is related to the adimensional moment of inertia parameter $\alpha_{\star}$ by the Darwin-Radau relation \citep{Murray}
\eqi\alpha_{\star}=\rp{2}{3}\left(1-\rp{2}{5}\sqrt{\rp{5}{2}\rp{q^{\star}_r}{f_{\star}}-1}\right).\lb{radau}\eqf
The obvious requirement that $\alpha_{\star}>0$ yields
\eqi f_{\star}>f^{\star}_{\rm min}=\rp{10}{29}q^{\star}_r=0.0130849,\eqf
while the the condition that $\alpha_{\star} < 2/5=0.4$, valid only for a homogeneous spherical body, implies\footnote{Note that \rfr{limite} automatically guarantees that the argument of the square root in \rfr{radau} is positive.}
\eqi f_{\star}< f^{\star}_{\rm max}=\rp{5}{4}q^{\star}_r=0.047326.\lb{limite}\eqf
As a reasonable figure for the star's flattening we take the average
\eqi f_{\star} =\rp{f^{\star}_{\rm min}+f^{\star}_{\rm max}}{2}=0.0302587,\eqf
which  yields
\eqi\alpha_{\star}=0.277011,\eqf
of the same order of magnitude of the evaluations by \citet{Ford} for main-sequence stars.
As a consequence, a realistic evaluation for the star's proper angular momentum is
\eqi S_{\star}=7.16483\times 10^{43}\ {\rm kg\ m^2\ s^{-1}}. \lb{spin}\eqf By comparison, the Sun's proper angular momentum, obtained from asteroseismology, amounts to \citep{Pijp1,Pijp2}
\eqi S_{\odot} = 1.9\times 10^{41}\ {\rm kg\ m^2\ s^{-1}},\eqf i.e., it is about 377 times smaller than \rfr{spin}.

In Table \ref{pianeta} we quote some relevant parameters of the planet.
\begin{table*}[t]
\caption{Relevant planet's parameters \citep{Wasp33}. We used $m_{\rm Jup}=0.00189864\times 10^{30}$ kg, $r_{\rm Jup}=71.492\times 10^6$ m. The planet's angular speed $\xi_{\rm p}$ has been assumed equal to the Keplerian mean motion $n=2\pi/P_{\rm b}$, where the orbital period is $P_{\rm b}=1.2198669\pm 0.0000012$ d \citep{Wasp33}. The values displayed for the rotational and tidal parameters $q^{\rm p}_r$ and $q^{\rm p}_t$ are lower bounds corresponding to \citep{Wasp33} $m^{\rm p}_{\rm max}=4.1\ m_{\rm Jup}$ for the planet's mass $m_{\rm p}$. The value of the relative semi-major axis used in $q_t^{\rm p}$ is $a=0.02555\pm 0.00017$ au \citep{Wasp33}. The figure for $j^{\rm p}_2$ has been obtained by using  $k_2^{\rm p}=0.3$ for the planet's Love number. The value used for the flattening $f_{\rm p}$ is the average of $f^{\rm p}_{\rm min}$ and $f^{\rm p}_{\rm max}$.}\label{pianeta}
\begin{tabular}{@{}lll}
\hline
Parameter (units) &  Numerical value  \\
\hline
$m_{\rm p}\ (m_{\rm Jup})$ & $<4.1$\\
$r_{\rm p}\ (r_{\rm Jup})$ & $1.497\pm 0.045$ \\
\hline
%
%
$\xi_{\rm p}$ (s$^{-1}$) & $5.96147\times 10^{-5}$ \\
$q^{\rm p}_r$ & $0.00838747$\\
$q^{\rm p}_t$ & $-0.0251564$\\
$j^{\rm p}_2$ & $2.09657\times 10^{-3}$ \\
$f_{\rm p}$ & $0.00668828$ \\
$\alpha_{\rm p}$ & $0.277011$ \\
$s_{\rm p}$ (kg m$^2$ s$^{-1}$) & $1.47243\times 10^{39}$\\
\hline
\end{tabular}
\end{table*}
We have assumed that the planet's angular speed $\xi_{\rm p}$ equals its orbital frequency $n$; for the planet's Love number  we assumed the moderate value\footnote{There is a wide variability in the possible values of $k_2^{\rm p}$ in hot Jupiters which can range from about 0.1 to 0.6 \citep{Boden}.} $k_2^{\rm p}=0.3$ \citep{Ragozzine}. Note that, contrary to the star case, the tidal parameter
\eqi q_t^{\rm p}\doteq-3\left(\rp{r_{\rm p}}{a}\right)^3\left(\rp{M_{\star}}{m_{\rm p}}\right)\eqf is about 3 times larger than the rotational parameter
\eqi q_r^{\rm p}\doteq\rp{\xi^2_{\rm p}r^3_{\rm p}}{Gm_{\rm p}},\eqf i.e. the tidal bulge raised on the planet by the star is larger than the centrifugal one due to its rotation. As a consequence, the tidal contribution to $j_2^{\rm p}$ dominates that due to the rotation; indeed,
\eqi \rp{k_2^{\rm p}}{3}q_r^{\rm p}=8.38747\times 10^{-4},\ -\rp{k_2^{\rm p}}{6}q_t^{\rm p}=1.25782\times 10^{-3}.\eqf Concerning the planet's proper angular momentum, repeating the same reasonings as for the star yields
\eqi s_{\rm p}=1.47243\times 10^{39}\ {\rm kg\ m^2\ s^{-1}},\eqf which is about 48000 times smaller than $S_{\star}$.
\section{The orbital effects}\lb{due}
Since the orbit of the planet is assumed circular \citep{Wasp33}, it is not possible to consider, not even in principle, either the periastron $\omega$ or the time of passage at periastron $t_p$. Thus,
we will not investigate the precessions induced in $\omega$ and the mean anomaly\footnote{Its precession is related to the change in $t_p$; see, e.g., \citet{Iorio2,Li}.} $\mathcal{M}$. Thus, let us consider the longitude of the ascending node $\Omega$ whose secular change is related to the precession of the orbital plane. Such an effect may impact some directly observable quantities in photometry of transiting extrasolar planets like the transit time duration  \citep{Miralda}. Preliminary numerical analyses have been performed by \citet{Ragozzine}.
\subsection{Classical and relativistic node precessions}\lb{clano}
Concerning the tidal effects accounted for by $q_t$, they do not secularly affect the node. Indeed, the tidal correction $U_{\rm tid}$ to the pointlike two-body Newtonian gravitational potential only contains $r^{-6}$ \citep{Ste,Ragozzine}, so that when it is averaged over one orbital revolution the angle
$\psi$ between the orbital angular momentum and the star/planet's spin axes, defined as $\psi=0$ when the two angular momenta are parallel and $\psi=\pi$ when they are antiparallel, does not appear in $\left\langle U_{\rm tid}\right\rangle$; actually, the node secular precession $\dot\Omega$ is proportional just to $\partial \left\langle U_{\rm tid}\right\rangle/\partial \psi$ according to the Lagrange planetary equations for the variation of the Keplerian orbital elements \citep{Murray}.

On the contrary, the centrifugal effects accounted for by $q_r$ do affect the nodes because the potential $U_{\rm rot}$ contains the angle $\psi$ \citep{Ragozzine}. The node precessions in the case of a two-body system with
arbitrary masses $m_{\rm A}$ and $m_{\rm B}$ and rotational quadrupole mass moments $q_r^{\rm A}$ and $q_r^{\rm B}$ have been worked out by \citet{LT1} and \citet{LT3}. In our case, the total precession is
\eqi \dot\Omega_{q_r}=\dot\Omega_{q_r}^{\star}+\dot\Omega_{q_r}^{\rm p},\eqf with
\begin{equation}
\begin{array}{lll}
\dot\Omega^{\star}_{q_r}&= &-\rp{3}{2}\left(\rp{2\pi}{P_{\rm b}}\right)\left(\rp{k^{\star}_2}{3}q^{\star}_r\right)\left(\rp{R_{\star}}{a}\right)^2\rp{\cos \Psi_{\star}}{(1-e^2)^2}\leq\\ \\
&\leq & 7.6138\times 10^{-10}\ {\rm s}^{-1},\\ \\
\dot\Omega^{\rm p}_{q_r}&=&-\rp{3}{2}\left(\rp{2\pi}{P_{\rm b}}\right)\left(\rp{k^{\rm p}_2}{3}q^{\rm p}_r\right)\left(\rp{r_{\rm p}}{a}\right)^2\rp{\cos \psi_{\rm p}}{(1-e^2)^2}=\\ \\
&=&(-5.880\times 10^{-11})\cos \psi_{\rm p}\ {\rm s}^{-1}.
\end{array}\lb{nodoprec}
 \end{equation}
 Note that $1\times 10^{-10}$ s$^{-1}$ corresponds to 1.78 deg d$^{-1}$, which is $9\times 10^9$ times larger than the node precession of Mercury induced by the Sun's oblateness \citep{Iorio05}.
Concerning the numerical value of $\dot\Omega^{\star}_{q_r}$ quoted in \rfr{nodoprec}, it has to be regarded as an upper limit computed for the largest possible value of $\cos\Psi_{\star}$. Actually, the angle $\Psi_{\star}$ should not be confused with\footnote{It is denoted as $\lambda$ by \citet{Wasp33} and \citet{hat2}.} $\Lambda_{\star}=251.2\pm 1.0$ deg \citep{Wasp33}, which is, instead,  the projected angle between the star's spin axis and the orbital angular momentum defined in such a way that $\Lambda_{\star}=0$ deg if the sky-projected angular momenta are parallel and $\Lambda_{\star}=\pi$ if they are antiparallel \citep{hat2}. In general, $\Lambda_{\star}$ can be derived from accurate radial-velocity data through the Rossiter-McLaughlin effect \citep{Ross}, but in the present case characterized by large stellar rotation its has been obtained from a line-profile analysis in a far more direct and less model-dependent way \citep{Wasp33}. The relation between $\Psi_{\star},I_{\star},\Lambda_{\star},i$ can be obtained as follows.
From the spherical law of cosines
\citep{Gel,Zwi}
\eqi\cos A=-\cos C\cos B+\sin C\sin B\cos a,\eqf   the identifications $A\rightarrow \Psi_{\star},B\rightarrow \pi-i,C\rightarrow I_{\star},a\rightarrow\Lambda_{\star}$ so that the segment $a$ is due to the plane of the sky, the segment $c$ is due to the orbital plane and the segment $b$ is the star's equatorial plane,
 yield \citep{hat2}
\eqi\cos\Psi_{\star}=\cos I_{\star}\cos i +\sin I_{\star}\sin i \cos\Lambda_{\star}.\lb{angol}\eqf
Given the determined values of $i$ and $\Lambda_{\star}$, \rfr{angol}, applied to the constraints on $I_{\star}$ of \rfr{minimoa}, tells us that the minimum and the maximum values for $\Psi_{\star}$ are
%
%
%
%
\begin{equation}
\begin{array}{lll}
I_{\star}&=&I_{\rm min}^{\star}\rightarrow\Psi_{\rm min}^{\star}=91.30\ {\rm deg},\\ \\
I_{\star}&=&97.20\ {\rm deg}\rightarrow\Psi_{\rm max}^{\star}=108.93\ {\rm deg},
\end{array}\lb{angoli}
 \end{equation}
 i.e. the orbit of WASP-33b can be considered  nearly polar.
 Consequently, we have
 %
%
%
%
 \begin{equation}
\begin{array}{lll}
\left|\cos\Psi_{\star}\right|_{\rm min}&=&0.022,\ \cos\Psi_{\star}<0,\\ \\
\left|\cos\Psi_{\star}\right|_{\rm max}&=&0.324,\ \cos\Psi_{\star}<0.
\end{array}\lb{coseniluce}
 \end{equation}
The value $I_{\star}=\tilde{I_{\star}}$  yields $\tilde{\Psi}_{\star}=108.78$ deg and $\cos\tilde{\Psi}_{\star}=-0.32$.
Note that the constraints on $\Psi_{\star}$ of \rfr{angoli} are compatible with the observations. Indeed, since the time series of the residual average spectral line profile of HD 15082 during the transits unambiguously reveals that the motion of the planet is retrograde \citep{Wasp33}, it must be $\cos\Psi_{\star}<0$.
Concerning $\psi_{\rm p}$, i.e. the angle between the planet's spin axis and the orbital angular momentum, it is not known and, at present, it seems that there is no way to measure it\footnote{(Collier Cameron A., private communication, June 2010).}. The largest possible value for $\dot\Omega_{q_r}^{\rm p}$ corresponds to $\psi_{\rm p}=0$ deg. Indeed, it could be naively  expected that the planet's rotation became
synchronized
and aligned with the orbital angular momentum  on a rather short timescale; there might be
a possibility that it gets into a Cassini state \citep{Hol}.

The  general relativistic gravitomagnetic (GM) spin-orbit effects in the case of a two-body system with
arbitrary masses $m_{\rm A}$ and $m_{\rm B}$ and spins $S_{\rm A}$ and $S_{\rm B}$ have been derived by \citet{LT1}, \citet{LT2}, \citet{LT3}. In our case,
by posing \eqi\nu\doteq \rp{m_{\rm p}}{M_{\star}}< 2.6\times 10^{-3},\eqf the total gravitomagnetic node precession is \eqi\dot\Omega_{\rm GM}=\dot\Omega^{\star}_{\rm GM}+\dot\Omega^{\rm p}_{\rm GM},\eqf with\footnote{In the test particle limit for one of the two bodies, we have the Lense-Thirring precession \citep{LT}. Actually, in our case $\dot\Omega_{\rm GM}^{\rm p}$ is negligible, and the Lense-Thirring  approximation is fully adequate for $\dot\Omega_{\rm GM}^{\star}$.}
\begin{equation}
\begin{array}{lll}
\dot\Omega^{\star}_{\rm GM} &=& \left(1+\rp{3}{4}\nu\right)\rp{2 G S_{\star}}{c^2 a^3(1-e^2)^{3/2}} =\\ \\
&=& 1.53\times 10^{-12}\ {\rm s^{-1}},\\\\
\dot\Omega^{\rm p}_{\rm GM}  &= & \left(1+\rp{3}{4\nu}\right)\rp{2 G s_{\rm p}}{c^2 a^3(1-e^2)^{3/2}}= 1\times 10^{-14}\ {\rm s^{-1}}.
\end{array}\lb{LTrates}
 \end{equation}
 Contrary to the previously examined precessions due to the oblateness, the gravitomagnetic node precessions do not depend on the angles between the orbital angular momentum and the bodies' spin axes.
 Note that $1\times 10^{-12}$ s$^{-1}$ corresponds to $651$ arcsec cty$^{-1}$, which is $3.25\times 10^5$ larger than the predicted solar Lense-Thirring  precession of the node of Mercury \citep{Iorio05}, and 15 times larger than the well-known gravitoelectric Einstein precession of the perihelion of Mercury in the field of the Sun.

Table \ref{effetti} summarizes the main precessional effects considered in this Section.
\begin{table*}[t]
\caption{Node precessions due to the classical centrifugal oblateness and the general relativistic gravitomagnetic fields of both the star and the planet.
The value of $\dot\Omega_{q_r}^{\star}$ corresponds to $|\cos\Psi_{\star}|_{\rm max}=0.324$.
 }\label{effetti}
\begin{tabular}{@{}llll}
\hline
$\dot\Omega_{q_r}^{\star}$ (s$^{-1}$) & $\dot\Omega_{q_r}^{\rm p}$ (s$^{-1}$) & $\dot\Omega_{\rm GM}^{\star}$ (s$^{-1}$) & $\dot\Omega_{\rm GM}^{\rm p}$ (s$^{-1}$) \\
\hline
$\leq 7.6138\times 10^{-10}$ & $(-5.880\times 10^{-11})\cos \psi_{\rm p}$ & $1.53\times 10^{-12}$ & $1\times 10^{-14}$\\
\hline
\end{tabular}
\end{table*}
The gravitomagnetic (star-induced) precession is $1-2$ orders of magnitude smaller than both the oblateness precessions. The largest one is due to the star's
oblateness; it overwhelms the planetary-induced precession by about 1 order of magnitude.
Concerning the uncertainties in the nominal values of the dominant precessions $\dot\Omega_{q_r}$, they can be conservatively evaluated by linearly propagating the errors in the parameters entering their expressions. It turns out
\begin{equation}
\begin{array}{lll}
\delta(\dot\Omega_{q_r}^{\star})&\leq & 0.106\ \dot\Omega_{q_r}^{\star}, \\ \\
\delta(\dot\Omega_{q_r}^{\rm p})&\leq & 0.163\ \dot\Omega_{q_r}^{\rm p};
\end{array}\lb{incertezza}
 \end{equation}
 the uncertainties are of the same order of, or larger by 1 order of magnitude than the general relativistic gravitomagnetic precessions. Actually, they may be larger because we have kept the Love numbers $k_2$ fixed to reference values.
 The previous considerations about the reciprocal orders of magnitude of the various node precessions hold for the upper bound of $\dot\Omega_{q_r}^{\star}$ due to $|\cos\Psi_{\star}|_{\rm max}$, but it is easy to realize that, substantially, retain their validity also for $|\cos\Psi_{\star}|_{\rm min}$ yielding $\dot\Omega_{q_r}^{\star}= 5.169\times 10^{-11}$ s$^{-1}$.

\subsection{The effect of a third body}
\begin{figure*}
\centerline{
\vbox{
\epsfysize= 5.0 cm\epsfbox{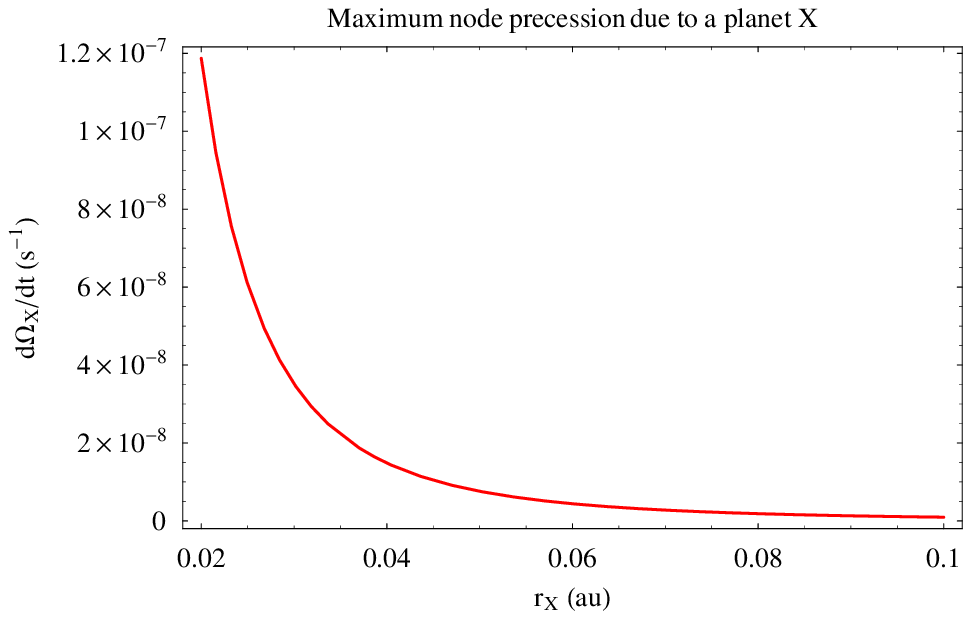}
\epsfysize= 5.0 cm\epsfbox{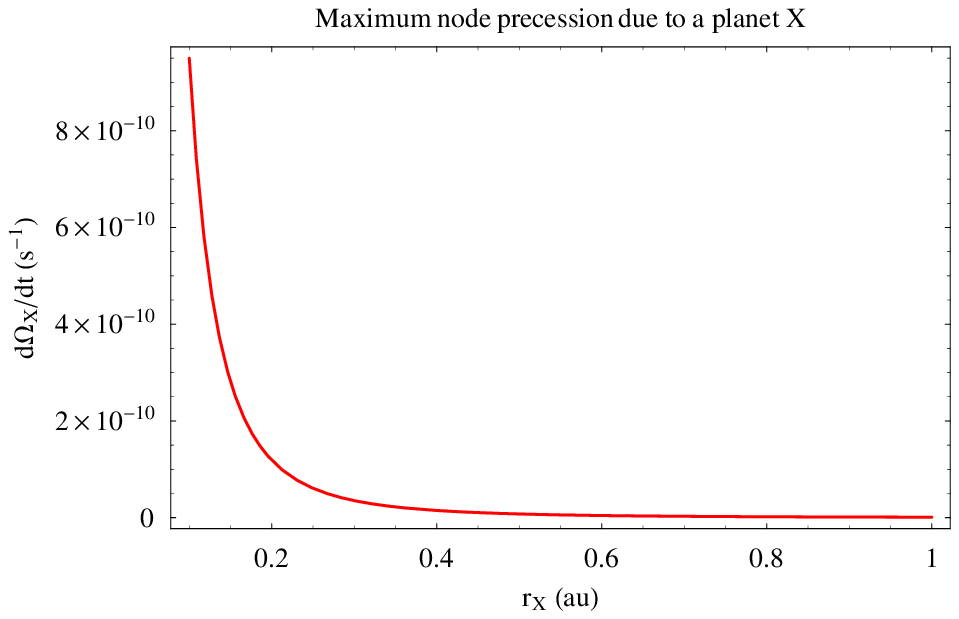}
\epsfysize= 5.0 cm\epsfbox{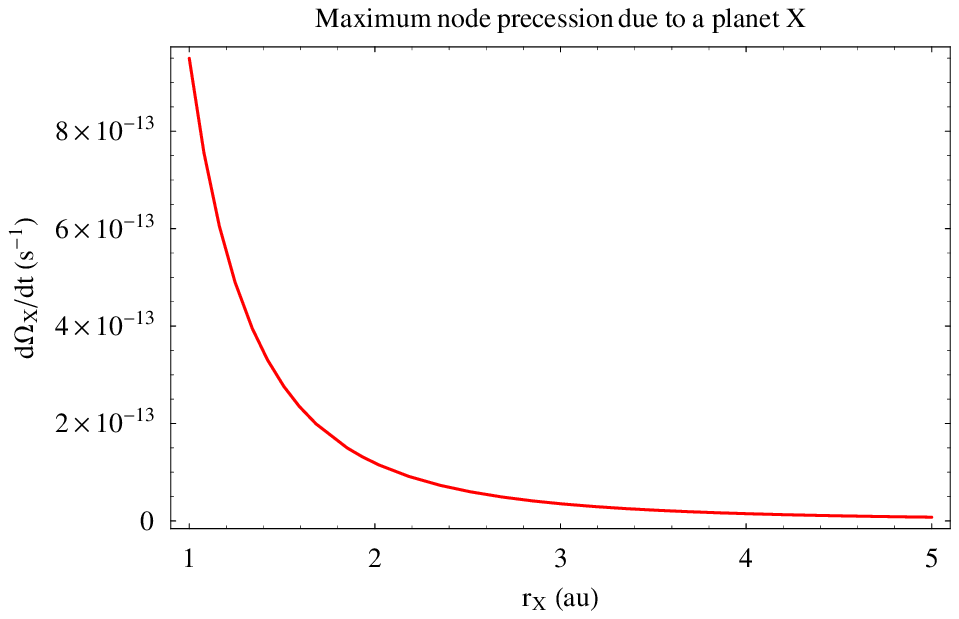}
\epsfysize= 5.0 cm\epsfbox{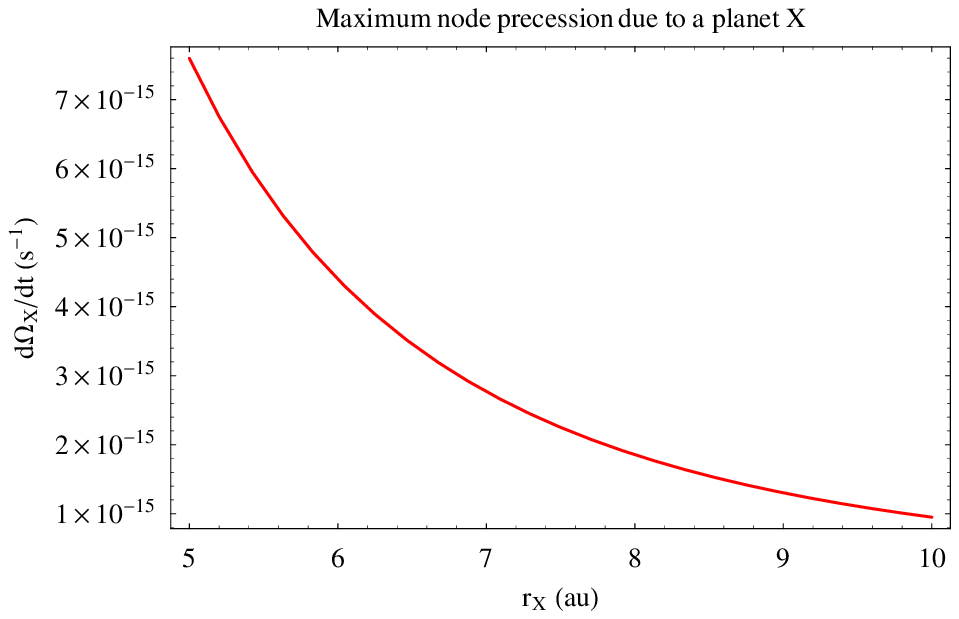}
      }
%
%
%
}
\caption{{Approximate upper bound $(3/2)(Gm_{\rm X}/r^3_{\rm X})(1/n)$ of the node precession of WASP-33b due to a putative planet X for $m_{\rm X}=m_{\rm Jup}$ as a function of its distance for $0.02\ {\rm au}\leq r_{\rm X}\leq 10$ au.}\label{tripla}}
\end{figure*}

The node precession induced by a distant third body $X$  can be computed quite generally by using the Lagrange planetary equations \citep{Murray} without making any a-priori assumptions on its location.
The perturbing potential induced by X is \citep{Ux}
\eqi U_{\rm X}=\rp{Gm_{\rm X}}{2r^3_{\rm X}}\left[r^2-3(\bds r\bds\cdot \bds{\hat{l}}_{\rm X})^2\right],\lb{UX}\eqf
where  $\bds{\hat{l}}_{\rm X}\doteq \bds{r}_{\rm X}/r_{\rm X}$ is a unit vector pointing towards X.
By denoting $l_x,l_y,l_z$ the direction cosines of $\bds{r}_X$, i.e. the components of  $\bds{\hat{l}}_{\rm X}$, we will express $U_{\rm X}$ as
\eqi U_{\rm X}=\rp{Gm_{\rm X}}{2r^3_{\rm X}}\left[r^2-3(xl_x+yl_y+zl_z)^2\right].\lb{UXnew}\eqf
In working out the secular effects by X on the orbit of WASP-33b, \rfr{UXnew} must be evaluated onto its unperturbed Keplerian ellipse and averaged over one orbital revolution; in doing that,  both $r_{\rm X}$ and $\bds{\hat{l}}_{\rm X}$ will be reasonably assumed constant over one orbital period of WASP-33b, consistently with the hypothesis that if X exists, it must be quite distant.
Useful formulas used in the calculation are
\begin{equation}
\begin{array}{lll}
x&=&r\left(\cos\Omega\cos u\ -\cos\Psi_{\star}\sin\Omega\sin u\right),\\\\
y&=&r\left(\sin\Omega\cos u + \cos\Psi_{\star}\cos\Omega\sin u\right),\\ \\
z&=& r\sin\Psi_{\star}\sin u,\\ \\
\cos f &=& \rp{\cos E-e}{1-e\cos E}, \\  \\
\sin f &=& \rp{\sqrt{1-e^2} \sin E}{1-e\cos E},\\ \\
r&=&a(1-e\cos E),\\ \\
dt &=&\rp{(1-e\cos E)}{n}dE,
\end{array}\lb{solite}
 \end{equation}
where $u\doteq\omega+f$ is the argument of latitude, $f$ is the true anomaly, and $E$ is the eccentric anomaly.
The Gauss equations for the variation of the Keplerian orbital elements \citep{Roy} straightforwardly yield
\begin{equation}
\begin{array}{lll}
\left\langle\left.\dert{\Omega}{t}\right|^{\rm (X)}\right\rangle &=& \rp{3 Gm_{\rm X}\csc\Psi_{\star}(1-e^2)^3}{2 r^3_{\rm X} n}\left[l_z\sin\Psi_{\star}+\right. \\ \\
&+&\left.\cos\Psi_{\star}\left(l_y\cos\Omega-l_x\sin\Omega\right)\right]\times\\ \\
&\times & \left[l_z\cos\Psi_{\star}+\sin\Psi_{\star}\times\right. \\ \\
&\times &\left.\left(l_x\sin\Omega-l_y\cos\Omega\right)\right].
\end{array}\lb{strazio2}
 \end{equation}
An upper bound on the magnitude of the node precession of WASP-33b due to a putative planet X can be naively obtained from
\eqi|\dot\Omega_{\rm X}|\lesssim \rp{3}{2}\left(\rp{Gm_{\rm X}}{r^3_{\rm X}}\right)\rp{1}{n}=\rp{3.18763\times 10^{21}\ {\rm m}^3\ {\rm s}^{-1}}{r^3_{\rm X}}\eqf for
\eqi m_{\rm X}=m_{\rm Jup}.\eqf
In Figure \ref{tripla} we plot it as a function of its distance $r_{\rm X}$ from 0.02 au to 10 au.
It shows that the effect of the star's oblateness would be mimicked by a planet of one jovian mass at about $r_{\rm X}\approx 0.1-0.15$ au, while an effect as large as the general relativistic gravitomagnetic precession would take place if X was at about 0.7 au. After 1 au, the impact of a jovian X would reduce to less than $10^{-12}$ s$^{-1}$, amounting to $10^{-15}$ s$^{-1}$ at $r_{\rm X}\approx 10$ au.
\subsection{Measurability of the nodal precession}\lb{gaaz}
Let us define as $x$ reference  direction the intersection between the plane of the sky, perpendicular to the line of sight, and the stellar equatorial plane, so that
\eqi \bds{\hat x}\doteq\rp{\bds{\hat{S}}_{\star}\bds\times\bds{\hat{\rho}}}{\sin I_{\star}}:\lb{vers1}\eqf in it $\bds{\hat{\rho}}$ is the unit vector of the line of sight, oriented towards the observer.
The line of nodes is the intersection of the orbital plane with the star's equatorial plane, and forms an angle $\Omega$ relative to the reference $x$ axis.  The unit vector of the line of the node is
\eqi \bds{ \hat{\tau} }\doteq\rp{ \bds{\hat{S}}_{\star} \bds\times  \bds{\hat{L}} }{\sin\Psi_{\star}},\lb{vers2}\eqf
where $\bds{\hat{L}}$ is the unit vector directed along the orbital angular momentum. From \rfr{vers1} and \rfr{vers2} it follows
\eqi\sin\Psi_{\star}\sin I_{\star}\cos\Omega
=(\bds{\hat{S}}_{\star} \bds\times \bds{\hat{L}})
\bds\cdot
(\bds{\hat{S}}_{\star} \bds\times \bds{\hat{\rho}}).\lb{megas}\eqf

Concerning the methods for measuring the node $\Omega$, we mention that spectroscopic variations of the star during the transits can, in principle, be used to measure the angle $\Omega$ of the line of nodes\footnote{Actually, the measurable quantity in \citet{Quel} is $\Omega\sin I_{\star}$.} \citep{Quel}. Recently, it has been pointed out by \citet{polar} that  the analysis of the polarization of the light scattered by the planets'atmospheres may allow to determine\footnote{With an ambiguity of $\pi$ \citep{polar}.}, among other things, $\Omega$ as well within a few percent, according to preliminary numerical investigations.
Moreover, it turns out that the\footnote{They are the Stokes coefficients defined by the Rayleigh scattering \citep{polar}.} $Q/F$ and $U/F$  polarized light curves are sensitive just to a time-variation of the node, as shown by Figure 4 of \citet{polar}. Instead, the total degree of polarization $P$ is not affected by $d\Omega/dt$.

In the following, we will mainly focus our attention  on the effects that a secular node variation has on the transit duration $ t_d $
for which relatively simple analytical expressions can be derived.
\subsection{The time variation of the transit duration}\lb{dura}
The precession of the orbital plane changes the duration of the transit. For a circular orbit arbitrarily inclined to the line of sight the transit duration can be written as
\eqi  t_d =\rp{2(R_{\star}+r_{\rm p})}{na}\cos\delta,\lb{trans}\eqf where $\delta$, the latitude of the transit on the stellar disk, is defined from
\footnote{While \citet{Deeg}-see his Figure 2-neglects the planet's radius, \citet{Miralda} takes it into account. Note that \citet{Miralda} uses the angle between the orbital plane and the line of sight, and denotes it as $\alpha$; in terms of our $i$, it is $\alpha=\pi/2+i$, so that $\sin\alpha\rightarrow \cos i$. Moreover,  \citet{Miralda} adopts the letter $\gamma$ for the latitude of the transit $\delta$.} \citep{Deeg,Miralda}
\eqi \sin\delta=\rp{a\cos i}{R_{\star}+r_{\rm p}}.\lb{trans2}\eqf Intuitively, if $i=90$ deg, the transit occurs along a diameter of the stellar disk, so that $\delta=0$, as confirmed by \rfr{trans2}; thus, \rfr{trans}  reduces to the intuitive form\footnote{In this case, the transit duration $ t_d $ is simply the time interval elapsed from the instant when the planet's disk touches the star's disk on one side of it to the instant when the planet's disk leaves the star's disk on the opposite side, thus traveling for a distance $2R_{\star}+2r_{\rm p}$ at a speed given by its orbital velocity $na$.  } \eqi  t_d =\rp{2(R_{\star}+r_{\rm p})}{na}.\eqf Instead, for $i=0$, no transit occurs; indeed, \rfr{trans2} loses its meaning  since it would imply $\sin \delta > 1$. The quantity $\sin\delta$ can be thought as an adimensional impact parameter, which is just one of the quantities determined by \citet{Wasp33}. By naming it $b$, they obtain
\eqi b = 0.155-0.218.\eqf
It should, now, be determined if the rate of change of $ t_d $ can be measured over a reasonable observing time baseline $\Delta t_{\rm obs}\approx 10$ yr. Since the typical values of the precession periods is quite longer than 10 yr, only the time derivative $d t_d /dt$ is relevant.
By differentiating \rfr{trans} with respect to\footnote{We do not take the derivative with respect to $a$ because the secular variation of the semimajor axis vanishes for all the dynamical effects considered.} $\delta$ it is possible to obtain
\eqi\rp{d t_d }{dt}=-\rp{2(R_{\star}+r_{\rm p})\sin\delta}{na}\left(\rp{d\delta}{dt}\right).\lb{deti}\eqf
From \rfr{trans2} it follows
\eqi\rp{d\delta}{dt}=-\rp{a\sin i}{(R_{\star}+r_{\rm p})\cos\delta}\left(\rp{di}{dt}\right),\lb{mazonga}\eqf which, substituted into \rfr{deti}, yields
\eqi \rp{d t_d }{dt}=\rp{2\tan\delta\sin i}{n}\left(\rp{di}{dt}\right).\lb{erra}\eqf
From the spherical law of cosines
\citep{Gel,Zwi}
\eqi\cos B=\sin C\sin A\cos b-\cos C\cos A\eqf  with the identifications\footnote{In such a way,  $\Omega$ results to be prograde with respect to the orbital motion, i.e. $\Omega$ follows it.} $A\rightarrow \Psi_{\star},B\rightarrow \pi-i,C\rightarrow I_{\star},b\rightarrow\pi-\Omega$,
 it turns out
 \footnote{\citet{Miralda} denotes $\beta$ the angle between his mean plane-which in our case coincides with the star's equator-and the line of sight, so that $\beta=\pi/2-I_{\star}$; with this change, eq. (9) and eq. (12) of \citet{Miralda} agrees with our \rfr{spherical} and \rfr{fina}. It must also be noted that eq. (4) by \citet{Miralda} tells us that his $i_p$ and $i$ coincide with our $\Psi_{\star}$.}
\eqi\cos i=\sin\Psi_{\star}\sin I_{\star}\cos\Omega+\cos\Psi_{\star}\cos I_{\star}.\lb{spherical}\eqf
From \rfr{spherical} it is possible to obtain $\Omega$ as a function of $I_{\star}$. A numerical investigation of it shows that $\Omega$ is close to 90 deg for the most likely values of $I_{\star}$.
 %
%
%
%
By differentiating \rfr{spherical} with respect to $\Omega$ one gets
\eqi \sin i\left(\rp{di}{dt}\right)=\sin\Psi_{\star}\sin I_{\star}\sin\Omega\left(\rp{d\Omega}{dt}\right).\lb{tasso}\eqf
Finally, \rfr{erra} with \rfr{tasso} yields
\eqi \rp{d t_d }{dt}=\rp{2\tan\delta\sin\Psi_{\star}\sin I_{\star}\sin\Omega}{n}\left(\rp{d\Omega}{dt}\right).\lb{fina}\eqf
To check intuitively the consistency of \rfr{spherical} and of its consequences, let us consider a simplified situation in which the line of sight, the orbital angular momentum and the star's spin axis are mutually orthogonal. For example, we could imagine to see the orbital plane vertically in front of us, aligned with the stellar spin axis in an exactly  polar configuration; actually, the real configuration of WASP-33b is not too far from the scenario described here. In this case, \rfr{spherical} tells us that $i=\Omega=90$ deg, as expected. Moreover, according to our expectations, \rfr{tasso} guarantees that the rate of change of $i$ does not vanish, being just equal to that of $\Omega$.
The magnitude of the time variation of the transit duration induced by the star's oblateness through the node precession is shown in Figure \ref{dtddt}.
\begin{figure}[ht!]
\includegraphics[width=\columnwidth]{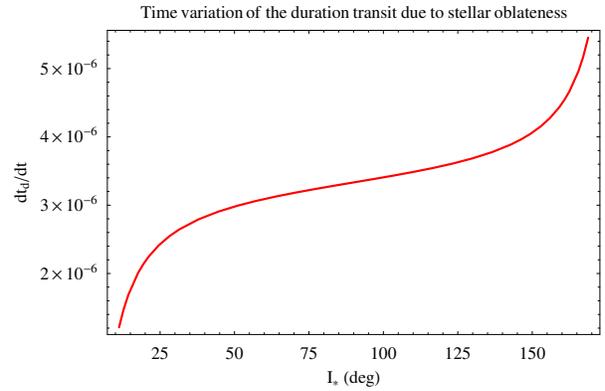}
 \caption{Time variation $d t_d /dt$ of the transit duration induced by the star's oblateness as a function of the angle $I_{\star}$ between the line of sight and the stellar spin axis for $11.22\ {\rm deg}<I_{\star}<168.77\ {\rm deg}$. }\lb{dtddt}
\end{figure}
It is of the order of \eqi\left.\rp{d t_d }{dt}\right|_{J_2^{\star}}=2-3\times 10^{-6}\eqf over all the allowed range of values for $I_{\star}$.
In Figure \ref{dtddtoj2} we plot $(d t_d /dt)\cos\psi_{\rm p}$ due to the node precession induced by the planet's centrifugal oblateness.
\begin{figure}[ht!]
\includegraphics[width=\columnwidth]{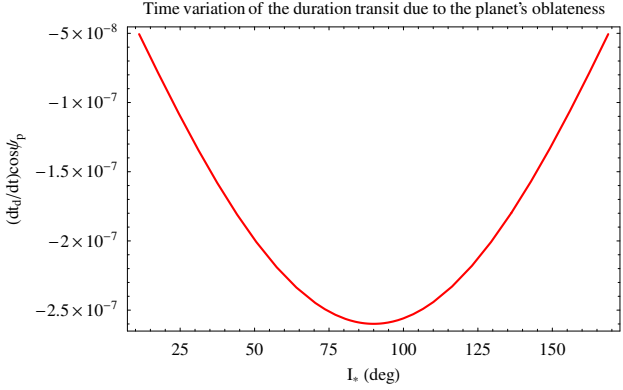}
 \caption{Time variation $(d t_d /dt)\cos\psi_{\rm p}$ of the transit duration induced by the planet's centrifugal oblateness as a function of the angle $I_{\star}$ between the line of sight and the stellar spin axis for $11.22\ {\rm deg}<I_{\star}<168.77\ {\rm deg}$. }\lb{dtddtoj2}
\end{figure}
It amounts to about $-2\times 10^{-7}\cos\psi_{\rm p}$ for the most likely values of $I_{\star}$.
Figure \ref{dtdtLT} depicts the effect of the gravitomagnetic node precession on the transit duration.
\begin{figure}[ht!]
\includegraphics[width=\columnwidth]{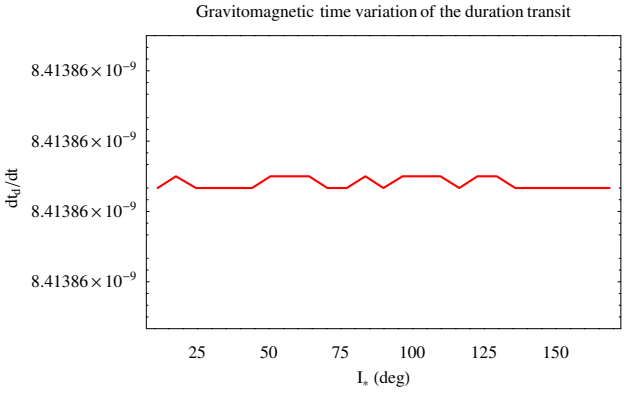}
 \caption{Time variation $d t_d /dt$ of the transit duration induced by the gravitomagnetic node precession as a function of the angle $I_{\star}$ between the line of sight and the stellar spin axis for $11.22\ {\rm deg}<I_{\star}<168.77\ {\rm deg}$. }\lb{dtdtLT}
\end{figure}
It is practically independent of $I_{\star}$, and amounts to
\eqi \left.\rp{d t_d }{dt}\right|_{\rm GM}=8.4\times 10^{-9}.\eqf Thus, it is about 360 times smaller than the variation induced by the stellar oblateness and 24 times smaller than the maximum value of transit duration variation due to the planet's centrifugal oblateness.

In order to evaluate the impact of a third body X on the variation of the transit duration, it is also necessary to derive \rfr{spherical} with respect to $\Psi_{\star}$. Indeed, contrary to the centrifugal oblateness and general relativity, X causes a non-vanishing precession of the angle between the orbital and the star's equatorial plane as well. The perturbative Gauss equations yield
\begin{equation}
\begin{array}{lll}
\left\langle\left.\dert{\Psi_{\star}}{t}\right|^{\rm (X)}\right\rangle &=& \rp{3 Gm_{\rm X}(1-e^2)^3}{2 r^3_{\rm X} n}\left(l_x\cos\Omega+l_y\sin\Omega\right)\times \\ \\
&\times &\left[l_z\cos\Psi_{\star}+\sin\Psi_{\star}\times\right. \\ \\
&\times &\left. \left(l_x\sin\Omega-l_y\cos\Omega\right)\right].
\end{array}\lb{strazio3}
 \end{equation}
 Note that its magnitude is of the same order of magnitude than the node precession of \rfr{strazio2}.
Thus, we have
\eqi
\begin{array}{lll}
\sin i \dert{i}{t} &=&\left(\sin\Psi_{\star}\cos I_{\star}-\cos\Psi_{\star}\sin I_{\star}\cos\Omega\right)\dert{\Psi_{\star}}{t}+\\ \\
&+&\sin\Psi_{\star}\sin I_{\star}\sin\Omega \dert{\Omega}{t},\lb{eho}
\end{array}
\eqf
so that
\eqi
\begin{array}{lll}
\dert{t_d}{t} &=& \rp{2\tan\delta}{n}\left[\left(\sin\Psi_{\star}\cos I_{\star}-\right.\right. \\ \\
&-&\left.\left.\cos\Psi_{\star}\sin I_{\star}\cos\Omega\right)\dert{\Psi_{\star}}{t} +\right. \\ \\
&+&\left.\sin\Psi_{\star}\sin I_{\star}\sin\Omega \dert{\Omega}{t}\right].\lb{ugaz}
\end{array}
\eqf
It is useful to look at the multiplicative trigonometric factors of $\dot\Psi_{\star}$ and $\dot\Omega$ in \rfr{ugaz}. A numerical investigation shows that the factor of $\dot\Omega$ ranges from $1000$ to $4500$ for the expected values of $I_{\star}$ (Figure \ref{trigonodo}), while the factor of $\dot\Psi_{\star}$ ranges from about $-4000$ to $+4000$ being zero close to $I_{\star}=\widetilde{I}_{\star}$ (Figure \ref{trigoincli}).
%
\begin{figure}[ht!]
\includegraphics[width=\columnwidth]{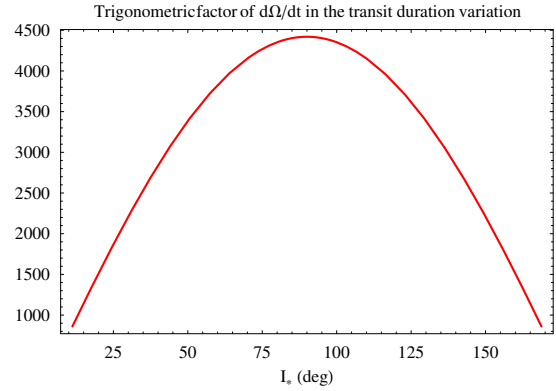}
 \caption{Trigonometric factor of $\dot\Omega$ in  $d t_d /dt$  as a function of the angle $I_{\star}$ between the line of sight and the stellar spin axis for $11.22\ {\rm deg}<I_{\star}<168.77\ {\rm deg}$. }\lb{trigonodo}
\end{figure}
\begin{figure}[ht!]
\includegraphics[width=\columnwidth]{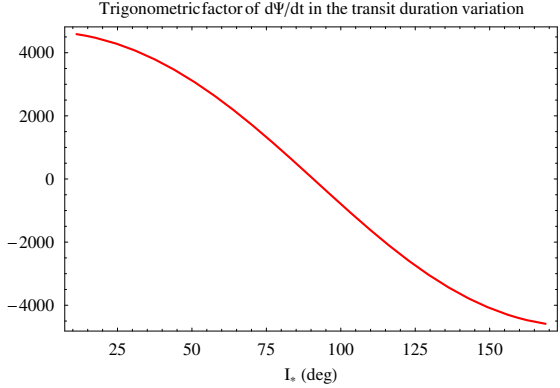}
 \caption{Trigonometric factor of $\dot\Psi_{\star}$ in  $d t_d /dt$  as a function of the angle $I_{\star}$ between the line of sight and the stellar spin axis for $11.22\ {\rm deg}<I_{\star}<168.77\ {\rm deg}$. }\lb{trigoincli}
\end{figure}

Concerning the accuracy with which a secular change of the transit duration may be measured over a typical time span $\Delta t=10$ yr, it seems reasonable to assume a $\approx 10^{-8}$ level\footnote{Collier Cameron A., private communication, June 2010.}; see also \citet{Miralda}, who speculates about the possibility of reaching a $\approx 10^{-9}$ level, and the thorough analysis by \citet{TRANS}. This would imply that the effect of the star's oblateness would be accurately measurable, while the detectability of  the variation in the transit duration due to the planet's centrifugal oblateness strongly depends on the angle $\psi_{\rm p}$ between the planet's spin axis and the orbital angular momentum. The general relativistic gravitomagnetic effect, instead, falls slightly below the measurability threshold. The same holds for a  third body X with $m_{\rm X}=m_{\rm Jup}$ located at about more than $0.7$ au.

\section{Summary and conclusions}\lb{quattro}
The quadrupole mass moment $J_2^{\star}$ and the proper angular momentum $S_{\star}$ of the fast rotating, main sequence star WASP-33 are 1900 and 400 times, respectively, larger than those of the Sun. Thus, huge classical and relativistic non-Keplerian orbital effects should affect the motion of the hot Jupiter harbored by WASP-33 which has been recently discovered with photometric transit measurements. If measurable, they would yield important information on the physical properties of the star and the planet.

In particular, the large inclination $\Psi_{\star}$ of the orbit of WASP-33b to the star's equator allows to consider the node precession $\dot\Omega$. It could be determined from the polarimetry of the light curves, and the time variation $d t_d /dt$ of the transit duration $ t_d $. The node rate of WASP-33b induced by $J_2^{\star}$ is $9\times 10^9$ times larger than the same effect for Mercury caused by the Sun's oblateness, while the general relativistic gravitomagnetic node precession-which is the only relativistic effect potentially measurable for WASP-33b if further analyses will confirm that its orbit is circular-is $3\times 10^5$ times larger than the Lense-Thirring effect for Mercury due to the Sun's rotation. We also considered the effect of the centrifugal oblateness  of the planet itself-the tidal bulge raised on it by the star does not affect the node-and of a putative distant third body X.
A conservative evaluation of the accuracy in measuring $d t_d /dt$ over 10 yr suggests a level $\approx 10^{-8}$.
The magnitudes of the induced time change in the transit duration are of the order of $3\times 10^{-6}, 2\times 10^{-7}, 8\times 10^{-9}$  for $J_2^{\star}$, the planet's rotational oblateness and general relativity, respectively. They depend on the angle $I_{\star}$ between the stellar spin axis and the line of sight which is, at present, unknown, and on the mass of the planet, for which only an upper bound is nowadays available. A yet undiscovered  planet X with the mass of Jupiter orbiting at at more than 1 au would induce a transit duration variation of less than $4\times 10^{-9}$.

%

In conclusion, it appears that the star's oblateness can be measured at a percent accuracy from a 10-yr analysis of the time variations of the duration transit. 
The signals from the planetary rotational oblateness are smaller by 1 order of magnitude, at the least; actually, they depend on the unknown angle $\psi_{\rm p}$ between the planet's equator and the orbital plane in such a way that they may turn out to be undetectable. The general relativistic gravitomagnetic signatures fall slightly below the measurability level. The presence in the system of a putative, distant third body, required in several theoretical schemes to explain the misalignment between the stellar and orbital angular momenta, seems to be undetectable if it has the mass of Jupiter and orbits at more than 1 au from the star.

The present analysis may be fruitfully repeated also if and when further planets moving along close inclined orbits around fast rotating stars will be discovered.
\section*{Acknowledgments}
I wish to thank A. Collier Cameron and J.N. Winn for useful information and remarks.


\begin{thebibliography}{}

\bibitem[\protect\citeauthoryear{Adams \& Laughlin}{2006a}]{Adamsa}
Adams F.C., Laughlin G., 2006a,
ApJ, 649,
992

\bibitem[\protect\citeauthoryear{Adams \& Laughlin}{2006b}]{Adamsb}
Adams F.C., Laughlin G., 2006b,
ApJ, 649,
1004

\bibitem[\protect\citeauthoryear{Adams \& Laughlin}{2006c}]{Adamsc}
Adams F.C., Laughlin G., 2006c, Int. J. Mod. Phys. D, 15, 2133

\bibitem[\protect\citeauthoryear{Anderson et al.}{2008}]{wasp5}
Anderson D.R., Gillon M., Hellier C., Maxted P.F.L., Pepe F., Queloz D., Wilson D.M., Collier Cameron A., Smalley B., Lister T.A., Bentley S.J., Blecha A., Christian D.J., Enoch B., Hebb L., Horne K., Irwin J., Joshi Y.C., Kane S.R., Marmier M., Mayor M., Parley N.R., Pollacco D.L., Pont F., Ryans R., Ségransan D., Skillen I., Street R.A., Udry S., West R.G., Wheatley P.J., 2008, MNRAS, 387, L4

\bibitem[\protect\citeauthoryear{Anderson et al.}{2010}]{Ross}
Anderson D.R., Hellier C., Gillon M., Triaud A.H.M.J., Smalley B., Hebb L., Collier Cameron A., Maxted P.F.L., Queloz D., West R.G., Bentley S.J., Enoch B., Horne K., Lister T.A., Mayor M., Parley N.R., Pepe F., Pollacco D., S\'{e}gransan D., Udry S., Wilson D.M., 2010, ApJ, 709, 159


\bibitem[\protect\citeauthoryear{Barker \& O'Connell}{1975}]{LT1}
Barker B.M.,  O'Connell R.F., 1975, Phys. Rev. D, 12,  329

\bibitem[\protect\citeauthoryear{Bodenheimer et al.}{2001}]{Boden}
Bodenhemer P., Lin D.N.C., Mardling R.A., 2001, ApJ, 548, 466


\bibitem[\protect\citeauthoryear{Carter et al.}{2008}]{TRANS}	
Carter J.A., Yee J.C., Eastman J., Gaudi B.S., Winn J.N., 2008, ApJ, 689, 499


\bibitem[\protect\citeauthoryear{Claret}{1995}]{Claret}
Claret A., 1995, Astronomy and Astrophysics Supplement, 114, 549

\bibitem[\protect\citeauthoryear{Collier Cameron et al.}{2007}]{wasp2}
Collier Cameron A., Bouchy F., H\'{e}brard G., Maxted P., Pollacco D., Pont F., Skillen I., Smalley B., Street R.A., West R.G., Wilson D.M., Aigrain S., Christian D.J., Clarkson W. I., Enoch B., Evans A., Fitzsimmons A., Fleenor M., Gillon M., Haswell C.A., Hebb L., Hellier C., Hodgkin S.T., Horne K., Irwin J., Kane S.R., Keenan F.P., Loeillet B., Lister T.A., Mayor M., Moutou C., Norton A.J., Osborne J., Parley N., Queloz D., Ryans R., Triaud A.H.M.J., Udry S., Wheatley P.J., 2007, MNRAS, 375, 951

\bibitem[\protect\citeauthoryear{Collier Cameron et al.}{2010}]{Wasp33}
Collier Cameron A., Guenther E., Smalley B., McDonald I., Hebb L., Andersen J., Augusteijn Th., Barros S.C.C., Brown D.J.A., Cochran W.D., Endl M., Fossey S.J., Hartmann M., Maxted P.F., Pollacco D., Skillen I., Telting J., Waldmann I.P., West R.G., 2010, MNRAS, 407, 507

\bibitem[\protect\citeauthoryear{Damour \& Sch\"{a}fer}{1988}]{LT2}
Damour T.,  Sch\"{a}fer G., 1988, Nuovo Cimento B, 101, 127

\bibitem[\protect\citeauthoryear{Deeg}{1998}]{Deeg}
Deeg H.-J., 1998, Photometric Detection of Extrasolar Planets by the Transit-Method. In: Rebolo R., Mart\'{\i}n E.L., Zapatero Osorio M.R., Brown dwarfs and extrasolar planets, Proceedings of a Workshop held in Puerto de la Cruz, Tenerife, Spain, 17-21 March 1997, ASP Conference Series vol. 134, pp. 216-223

\bibitem[\protect\citeauthoryear{Fluri \& Berdyugina}{2010}]{polar}
Fluri D.M., Berdyugina S.V., 2010, A\& A, 512, A59

\bibitem[\protect\citeauthoryear{Ford et al.}{1999}]{Ford}
Ford E., Rasio F.A., Sills A., 1999, ApJ, 514, 411

\bibitem[\protect\citeauthoryear{Gellert et al.}{1989}]{Gel}
Gellert W., Gottwald S., Hellwich M., K\"{a}stner H., K\"{u}nstner H. (eds.), 1989, Spherical Trigonometry. $\S$12 in VNR Concise Encyclopedia of Mathematics, 2nd ed.  (New York: Van Nostrand Reinhold), pp. 261-282

\bibitem[\protect\citeauthoryear{Gillon et al.}{2009}]{wasp5b}
Gillon M., Smalley B., Hebb L., Anderson D.R., Triaud A.H.M.J., Hellier C., Maxted P.F.L., Queloz D., Wilson D.M., 2009, A\&A, 496 , 259

\bibitem[\protect\citeauthoryear{Gizon \& Solanki}{2003}]{Gizon}
Gizon L., Solanki S.K., 2003, ApJ, 655, 589, 1009

\bibitem[\protect\citeauthoryear{Grenier et al.}{1999}]{Gren}
Grenier S., Baylac M.-O., Rolland L. Burnage R., Arenou F., Briot D., Delmas F., Duflot M., Genty V., G\'{o}mez A.E., Halbwachs J.-L., Marouard M., Oblak E., Sellier A., 1999, Astronomy and Astrophysics Supplement, 137, 451

\bibitem[\protect\citeauthoryear{Heyl \& Gladman}{2007}]{Heyl}
Heyl J.S., Gladman B.J., 2007, MNRAS, 377, 1511

\bibitem[\protect\citeauthoryear{Henry \& Winn}{2008}]{Henry}
Henry G.W., Winn J.N., 2008, AJ, 135, 68

\bibitem[\protect\citeauthoryear{Heyl \& Gladman}{2007}]{Heyl}
Heyl J.S.,  Gladman B.J. 2007, MNRAS, 377, 1511

\bibitem[\protect\citeauthoryear{Hogg et al.}{1991}]{Ux}
Hogg D., Quinlan G., Tremaine S., 1991, AJ, 101, 2274

\bibitem[\protect\citeauthoryear{Iorio}{2005}]{Iorio05}
Iorio L., 2005, A\&A, 431, 385

\bibitem[\protect\citeauthoryear{Iorio}{2006}]{Iorio}
Iorio L., 2006, New Astronomy, 11, 490

\bibitem[\protect\citeauthoryear{Iorio}{2007}]{Iorio2}
Iorio L., 2007, ApSS, 312, 331

\bibitem[\protect\citeauthoryear{Jord\'{a}n \& Bakos}{2008}]{Jordan}
Jord\'{a}n A., Bakos G.\'{A}., 2008,
 ApJ, 685,
 543


\bibitem[\protect\citeauthoryear{Laskar \& Gastineau}{2009}]{Lask}	
Laskar J., Gastineau M., 2009, Nature, 459, 817

\bibitem[\protect\citeauthoryear{Lense \& Thirring}{1918}]{LT}
Lense J., Thirring H., 1918,
Phys. Z.,
19,
156

\bibitem[\protect\citeauthoryear{Li}{2010}]{Li}
Li L.-S., 2010, ApSS, 327, 59


\bibitem[\protect\citeauthoryear{Mashhoon}{2007}]{Mash}
Mashhoon B., Gravitoelectromagnetism: A Brief Review, 2007, in: Iorio L. (ed.) The Measurement of Gravitomagnetism: A Challenging Enterprise, (Hauppauge: Nova Publishers), pp. 29-39


\bibitem[\protect\citeauthoryear{Miralda-Escud\'{e}}{2002}]{Miralda}
Miralda-Escud\'{e} J. 2002, ApJ, 564, 1019

\bibitem[\protect\citeauthoryear{Murray \& Dermott}{2000}]{Murray}
Murray C.D.,  Dermott S.F., 2000, Solar System Dynamics (New York:
Cambridge Univ. Press)

\bibitem[\protect\citeauthoryear{Narita et al.}{2009}]{hat1}
Narita N., Sato B., Hirano T., Tamura M., 2009, PASJ, 60, L35

\bibitem[\protect\citeauthoryear{P\'{a}l \& Kocsis}{2008}]{Pal}
P\'{a}l A., Kocsis B., 2008,
MNRAS, 389,
191

\bibitem[\protect\citeauthoryear{Perryman et al.}{1997}]{Perry}
Perryman M.A.C., Lindegren L., Kovalevsky J., Hoeg E., Bastian U., Bernacca P.L., Cr\'{e}z\'{e} M., Donati F., Grenon M., van Leeuwen F., van der Marel H., Mignard F., Murray C.A., Le Poole R.S., Schrijver H., Turon C., Arenou F., Froeschl\'{e} M., Petersen C.S., 1997, A\&A, 323, L49

\bibitem[\protect\citeauthoryear{Pijpers}{1998}]{Pijp1}
Pijpers F.P., 1998, MNRAS, 297, L76

\bibitem[\protect\citeauthoryear{Pijpers}{2003}]{Pijp2}
Pijpers F.P., 2003,
A\& A, 402, 683

\bibitem[\protect\citeauthoryear{Pireaux et al.}{2007}]{J2Sun}	
Pireaux S., Standish E.M., Pitjeva E.V.,  Rozelot J.-P., 2007, 	Solar quadrupole moment from planetary ephemerides: present state of the art, in: Proceedings of the International Astronomical Union (2006), 2, 473. (Cambridge: Cambridge Univ. Press)

\bibitem[\protect\citeauthoryear{Queloz et al.}{2000}]{Quel}
Queloz D., Eggenberger A., Mayor M., Perrier C., Beuzit J.L., Naef D., Sivan J.P., Udry S., 2000, A\&A, 359, L13

\bibitem[\protect\citeauthoryear{Queloz et al.}{2010}]{wasp8}
Queloz D.,  Anderson D.,  Collier Cameron A.,  Gillon M.,  Hebb L.,  Hellier C.,  Maxted P.,  Pepe F.,  Pollacco D.,  Segransan D.,  Smalley B.,  Udry S., R West R., 2010, A\&A, submitted

\bibitem[\protect\citeauthoryear{Ragozzine \& Wolf}{2009}]{Ragozzine}
Ragozzine D., Wolf A.S., 2009, ApJ, 698, 1778

\bibitem[\protect\citeauthoryear{Roy}{2005}]{Roy}
Roy A.E., 2005, Orbital Motion. Fourth Edition. (Bristol: Institute of Physics)

\bibitem[\protect\citeauthoryear{Sterne}{1939}]{Ste}
Sterne T.E., 1939, MNRAS, 99, 451

\bibitem[\protect\citeauthoryear{West et al.}{2009}]{wasp15}
West R.G., Anderson D.R., Gillon M., Hebb L., Hellier C., Maxted P.F.L., Queloz D., Smalley B., Triaud A.H.M.J., Wilson D.M., Bentley S.J., Collier Cameron A., Enoch B., Horne K., Irwin J., Lister T.A., Mayor M., Parley N., Pepe F., Pollacco D., Segransan D., Spano M., Udry S., Wheatley P.J., AJ, 137, 4834

\bibitem[\protect\citeauthoryear{Wex}{1995}]{LT3}
 Wex N., 1995, Class. Quant. Grav., 12, 983

 \bibitem[\protect\citeauthoryear{Winn \& Holman}{2005}]{Hol}
Winn J.N., Holman M.J., 2005, ApJ, 628, L159

\bibitem[\protect\citeauthoryear{Winn et al.}{2009}]{hat2}
Winn J.N., Johnson J.A., Albrecht S., Howard A.W., Marcy G.W., Crossfield I.J., Holman M.J., 2009, ApJ, 703, L99

\bibitem[\protect\citeauthoryear{Zwillinger}{1995}]{Zwi}
Zwillinger~D.~(ed.), 1995, Spherical Geometry and Tri\-go\-no\-me\-try. $\S$6.4 in CRC Standard Mathematical Tables and Formulae.  (Boca Raton, FL: CRC Press), pp. 468-471

\end{thebibliography}
\end{document}